\documentclass[conference]{sigcomm-alternate}
\usepackage{graphicx,latexsym,epsfig,amssymb,amsmath,subfigure}

\begin{document}
%\pagenumbering{roman}

\title{On the Economics of Cloud Markets}

\numberofauthors{2}
\author{
        \alignauthor
	Ranjan Pal\\
	\affaddr{Department of Computer Science}\\
	\affaddr{University of Southern California, USA}\\
	\affaddr{rpal@usc.edu}\and
    \alignauthor
    Pan Hui \\
    \affaddr{Deutsche Telekom Laboratories}\\
    \affaddr{Berlin, Germany}\\
    \affaddr{Pan.Hui@telekom.de}
}
\maketitle
%\small
\begin{abstract}
Cloud computing is a paradigm that has the potential to transform and revolutionalize the next generation IT industry by making software available to end-users as a service. A cloud, also commonly known as a cloud network, typically comprises of hardware (network of servers) and a collection of softwares that is made available to end-users in a \emph{pay-as-you-go} manner. Multiple public cloud providers (ex., Amazon) co-existing in a cloud computing market provide similar services (software as a service) to its clients, both in terms of the nature of an application, as well as in quality of service (QoS) provision. The decision of whether a cloud hosts (or finds it profitable to host) a service in the long-term would depend jointly on the price it sets, the QoS guarantees it provides to its customers , and the satisfaction of the advertised guarantees. In this paper, we devise and analyze three \emph{inter-organizational} economic models relevant to cloud networks. We formulate our problems as \emph{non co-operative} price and QoS games between \emph{multiple} cloud providers existing in a cloud market. We prove that a \emph{unique} pure strategy Nash equilibrium (NE) exists in two of the three models. Our analysis paves the path for each cloud provider to 1) know what prices and QoS level to set for end-users of a given service type, such that the provider could exist in the cloud market, and 2) practically and dynamically provision appropriate capacity for satisfying advertised QoS guarantees. 
\end{abstract}
% for peerreview papers, inserts a page break and creates the second title.
% Will be ignored for other modes.
%\IEEEpeerreviewmaketitle

\section{Introduction}
Cloud computing is a type of Internet-based computing, where shared resources, hardware, software, and information are provided to end-users in an \emph{on demand} fashion. It is a paradigm that has the potential to transform and revolutionalize the IT industry by making software available to end-users as a service \cite{katz}. A public cloud typically comprises of hardware (network of servers) and a collection of softwares that is made available to the general public in a \emph{pay-as-you-go} manner. Typical examples of companies providing public clouds include \emph{Amazon, Google, Microsoft, E-Bay,} and commercial banks. %Some examples of cloud computing services include Google Docs and Amazon Web Services. 
Public cloud providers usually provide Software as a Service (SaaS),  Platform as a Service (PaaS), and Infrastructure as a Service (IaaS).The advantage of making software available as a service is three-fold \cite{katz}, 1) the service providers benefit from simplified software installation, maintenance, and centralized versioning, 2) end-users can access the software in an `anytime anywhere' manner, can store data safely in the cloud infrastructure, and do not have to think about provisioning any hardware resource due to the illusion of infinite computing resources available on demand, and 3) end-users can pay for using computing resources on a short-term basis (ex., by the hour or by the day) and can release the resources on task completion. Similar benefit types are also obtained by making both, platform as well as infrastructure available as service. 

Cloud economics will play a vital role in shaping the cloud computing industry of the future. In a recent Microsoft white paper titled ``Economics of the Cloud", it has been stated that the computing industry is moving towards the cloud driven by three important economies of scale: 1) large data centers can deploy computational resources at significantly lower costs than smaller ones, 2) demand pooling improves utilization of resources, and 3) multi-tenancy lowers application maintenance labor costs for large public clouds. The cloud also provides an opportunity to IT professionals to focus more on technological innovation rather than thinking of the budget of "keeping the lights on". The economics of the cloud can be thought of having two dimensions: 1) intra-organization economics and 2) inter-organization economics. Intra-organization economics deals with the economics of internal factors of an organization like labor, power, hardware, security, etc., whereas inter-organization economics refers to the economics of market competition factors between organizations. Examples of some popular factors are price, QoS, reputation, and customer service. In this paper, we focus on inter-organizational economic issues.  

\begin{figure}
\centering
\includegraphics[width=3.3in]{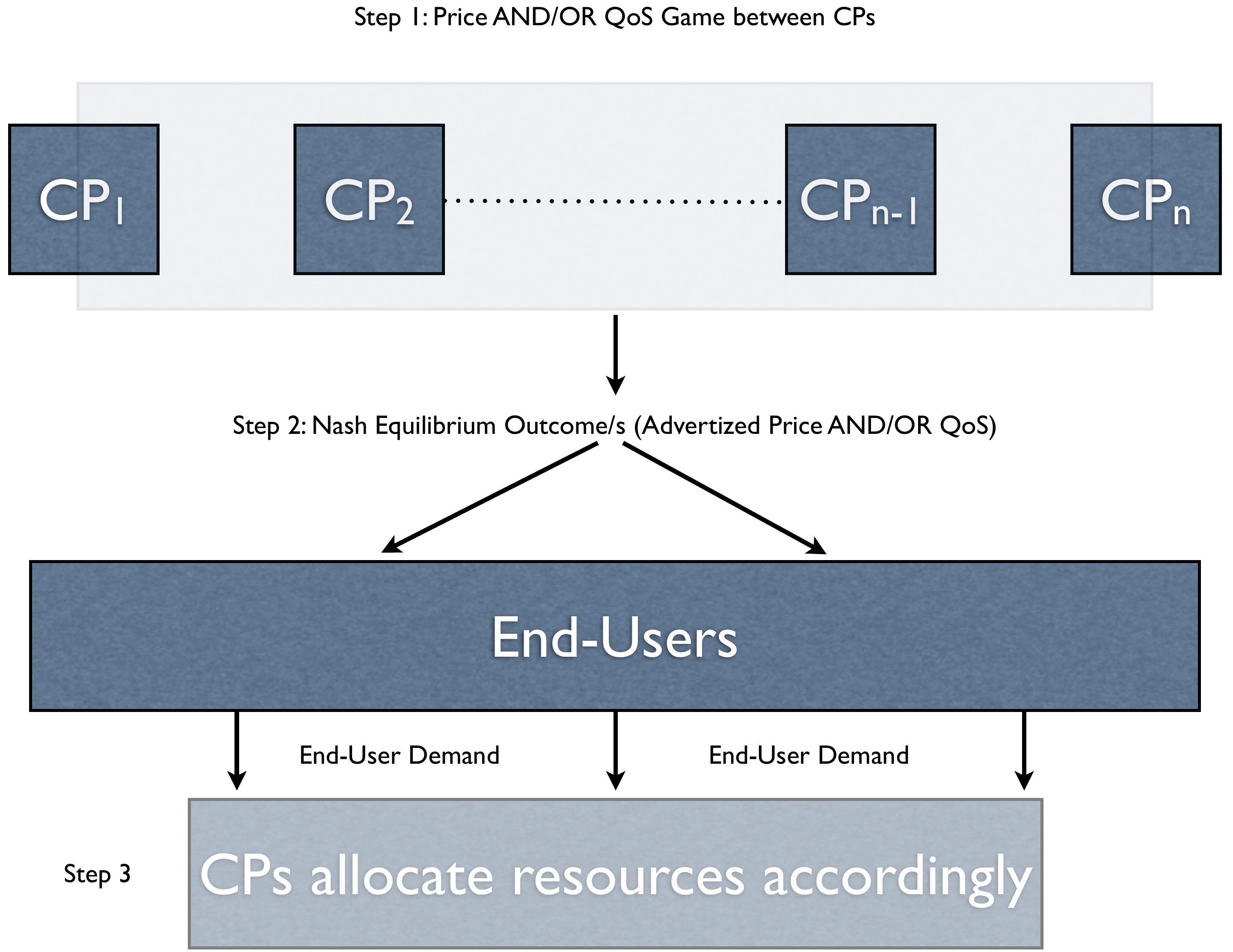}
\caption{Pricing Directed Resource Allocation in Clouds.\label{figure1}}
\end{figure}

Multiple public cloud providers (ex., \emph{Amazon, Google, Microsoft,} etc.,) co-existing in a cloud computing market provide similar services (software as a service, ex., \emph{Google Docs} and \emph{Microsoft Office Live}) to its clients, both in terms of the nature of an application, as well as in quality of service (QoS) provision. The decision of whether a cloud hosts (or finds it profitable to host) a service in the long-term would (amongst other factors)  depend jointly on the price it sets, the QoS guarantees it provides to its customers\footnote{A cloud provider generally gets requests from a cloud customer, which in turn accepts requests from Internet end-users. Thus, typically, the clients/customers of a cloud provider are the cloud customers. However, for modeling purposes, end-users could also be treated as customers. (See Section 2)}, and the satisfaction of the advertised guarantees. Setting high prices might result in a drop in demand for a particular service, whereas setting low prices might attract customers at the expense of lowering cloud provider profits. Similarly, advertising and satisfying high QoS levels would favor a cloud provider (CP) in attracting more customers. The price and QoS levels set by the CPs thus drive the end-user demand, which, apart from determining the market power of a CP also plays a major role in CPs estimating the minimal resource capacity to meet their advertised guarantees (Figure~\ref{figure1}). The estimation problem is an important challenge in cloud computing with respect to resource provisioning because a successful estimation would prevent CPs to provision for the peak, thereby reducing resource wastage. 

The competition in prices and QoS amongst the cloud providers entails the formation of non-cooperative games amongst competitive CPs. Thus, we have a \emph{distributed system} of CPs (players in the game), where each CP wants to maximize its own profits and would tend towards playing a Nash equilibrium\footnote{A group of players is in Nash equilibrium if each one is making the best decision(strategy) that he or she can, taking into account the decisions of the others.} (NE) strategy (i.e., each CP would want to set the NE prices and QoS levels.), whereby the whole system of CPs would have no incentive to deviate from the Nash equilibrium point, i.e., the vector of NE strategies of each CP. However, for each CP to play a NE strategy, the latter should mathematically exist. In this paper, we address the important problem of Nash Equilibrium characterization of \emph{different types} of price and QoS games relevant to cloud networks, its properties, practical implementability (convergence issues), and the sensitivity analysis of NE price/QoS variations by any CP on the price and QoS levels of other CPs. Our problem is important from a resource provisioning perspective as mentioned in the previous paragraph, apart from it having obvious strategic importance on CPs in terms of sustenance in the cloud market.

In regard to market competition driven network pricing, there exists research work in the domain of multiple ISP interaction and tiered Internet services \cite{llui}\cite{ssrt}, as well as in the area of resource allocation and Internet congestion management \cite{hccr}\cite{jpw}\cite{mvr}. However, the market competition in our work relates to optimal capacity planning and resource provisioning in clouds. There is the seminal work by Songhurst and Kelly \cite{sk} on pricing schemes based on QoS requirements of users. Their work address multi-service scenarios and derive pricing schemes for each service based on the QoS requirements for each, and in turn bandwidth reservations. This work resembles ours to some extent in the sense that the price and QoS determined can determine optimal bandwidth provisions. However, it does not account for market competition between multiple providers and only focus on a single service provider providing multiple services, i.e., the paper addresses an intra-organization economics problem. However, in this paper, we assume single-service scenarios by multiple service providers. 

Our proposed theory analyzes a few basic inter-organizational economic models through which cloud services \emph{could} be priced under market competition. The evolution of commercial public cloud service markets is still in its inception. However, with the gaining popularity of cloud services, we expect a big surge in public cloud services competition in the years to come. The models proposed in this paper take a substantial step in highlighting relevant models to the cloud networking community for them adopt so as to appropriately price current and future cloud services. In practice, scenarios of price and/or QoS competition between organizations exist in the mobile network services and ISP markets. For example, AT\&T and Verizon are competing on service, i.e., Verizon promises to provide better coverage to mobile users than AT\&T, thereby increasing its propensity to attract more customers. Similarly, price competition between ISPs always existed for providing broadband services at a certain given bandwidth guarantee. Regarding our work, we also want to emphasize 1) we do not make any claims about our models being the only way to model inter-organizational cloud economics\footnote{We only model price and QoS as parameters. One could choose other parameters (in addition to price and QoS, which are essential parameters) and a different analysis mechanism than ours to arrive at a different model.} and 2) there is a dependency between intra-organizational and inter-organizational economic factors, which we do not account in this paper due to modeling simplicity. However, through our work, we definitely provide readers with a concrete modeling intuition to go about addressing problems in cloud economics. \emph{To the best of our knowledge, we are the first to provide an analytical model on inter-organizational cloud economics.} 

\emph{Our Contributions}
\begin{enumerate}
\item We formulate a \emph{separable end-user demand function} for each cloud provider w.r.t. to price and QoS levels set by them and derive their individual utility functions (profit function). We then define the various price-QoS games that we analyze in the paper. (See Section 2)
\vspace{-3mm}
\item We develop a model where the QoS guarantees provided by public CPs to end-users for a particular application type are pre-specified and fixed, and the cloud providers compete for prices. We formulate a non-cooperative price game amongst the players (i.e., the cloud providers) and prove that there exists a unique Nash equilibrium of the game, and that the NE could be practically computed (i.e., it converges). (See Section 3)
\vspace{-3mm}
\item We develop a non-cooperative game-theoretic model where public cloud providers \emph{jointly} compete for the price and QoS levels related to a particular application type. We show the existence and convergence of Nash equilibria (See Section 4). As a special case of this model, we also analyze the case where prices charged to Internet end-users are pre-specified and fixed, and the cloud providers compete for QoS guarantees only. The models mentioned in contributions 3 and 4 drive optimal capacity planning and resource provisioning in clouds, apart from maximizing CP profits. (See Section 4)
\vspace{-3mm}
\item We conduct a sensitivity analysis on various parameters of our proposed models, and study the effect of changes in the parameters on the equilibrium price and QoS levels of the CPs existing in a cloud market. Through a sensitivity analysis, we infer the effect of price and QoS changes of cloud providers on their respective profits, as well as the profits of competing CPs. (See Sections 3 and 4)\footnote{We study Nash equilibrium convergence as its proves the achievability of an equilibrium point in the market. We emphasize here that the existence of Nash equilibrium does not imply achievability as it may take the cloud market an eternity to reach equilibrium, even though there may exist one theoretically.}
\end{enumerate}
\section{Problem Setup}
We consider a market of $n$ competing cloud providers, where each provider services application types to end-users at a given QoS guarantee. We assume that end-users are customers of cloud providers in an \emph{indirect} manner, i.e., Internet end-users use online softwares developed by companies (cloud customers), that depend on cloud providers to service their customer requests. Each CP is in competition with others in the market for services provided on the \emph{same type} of application w.r.t functionality and QoS guarantees. For example, Microsoft and Google might both serve a word processing application to end-users by providing similar QoS guarantees. Here, the word processing application represents a particular `type'. For a given application type, we assume that each end user signs a contract with a particular CP for a given time period\footnote{In this paper, the term `time-period' refers to the time duration of a contract between the CP and end-users.}, and within that period it does not switch to any other CP for getting service on the same application type. Regarding contracts between a CP and its end-users, we assume that a cloud customer forwards service requests to a cloud provider on behalf of end-users, who sign up with a cloud customer (CC) for service. The CP charges its cloud customer, who is turn charges its end-users (Figure 2). We approximate this two-step charging scheme by modeling a virtual one-step scheme, where a CP charges end-users directly\footnote{We assume here that prices are negotiated between the CP, CC, and end-users and there is a virtual direct price charging connection between the CP and its end-users. We make this approximation for modeling simplicity.}.  

\begin{figure}
\centering
\includegraphics[width=3.3in]{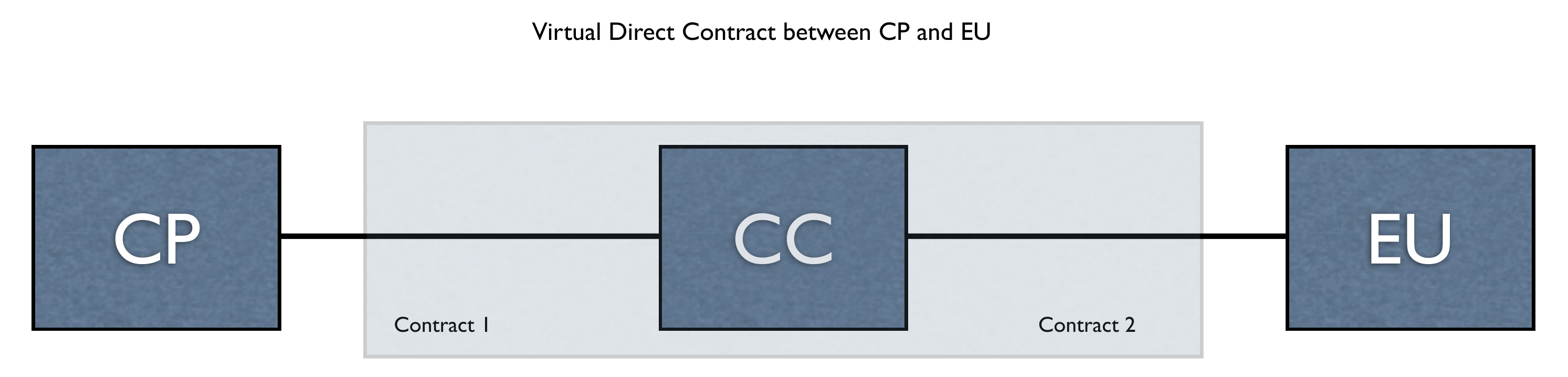}
\caption{Virtual Direct Contract between CP and EU.\label{figure2}}
\end{figure} 

In a given time period, each CP $i$ positions itself in the market by selecting a price $p_{i}$ and a QoS level $s_{i}$ related to a given application type. Throughout the paper, we assume that the CPs compete on a single given type\footnote{In reality, each CP may in general service several application types \emph{concurrently}. We do not model this case in our paper and leave it for future work. The case for single application types gives interesting results, which would prove to be useful in analyzing the multiple concurrent application type scenario.}. We define $s_{i}$ as the difference between a benchmark response time upper bound, $\overline{rt}$, and the actual response time $rt_{i}$, i.e., $s_{i} = \overline{rt} - rt_{i}$. For example, if for a particular application type, every CP would respond to an end-user request within 10 seconds, $\overline{rt} = 10$. The response time $rt_{i}$ may be defined, either in terms of the expected steady state response time, i.e., $rt_{i} = E(RT_{i})$, or in terms of $\phi$-percentile performance, $rt_{i}(\phi)$, where $0 < \phi < 1$. Thus, in terms of $\phi$-percentile performance\footnote{As an example, in cloud networks we often associate provisioning power according to the 95th percentile use. Likewise, we could also provision service capacity by accounting for percentile response time guarantees.}, $P(RT_{i} < rt_{i}(\phi)) = \phi$.

We model each CP $i$ as an M/M/1 queueing system, where end-user requests arrive as a Poisson process with mean rate $\lambda_{i}$, and gets serviced at a rate $\mu_{i}$. We adopt an M/M/1 queueing system because of three reasons: 1) queueing theory has been traditionally used in request arrival and service problems, 2) for our problem, assuming an M/M/1 queueing system ensures tractable analyses procedures that entails deriving nice closed form expressions and helps understand system insights in a non-complex manner, without sacrificing a great deal in capturing the real dynamics of the actual arrival-departure process, and 3) The Markovian nature of the service process helps us generalize expected steady state analysis and percentile analysis together. According to the theory of M/M/1 queues, we have the following standard results \cite{bg}.
\begin{equation}
rt_{i} = \frac{1}{\mu_{i} - \lambda_{i}},
\end{equation}

\begin{equation}
rt_{i}(\phi) = \frac{ln(\frac{1}{1 - \phi})}{\mu_{i}(\phi) - \lambda_{i}},
\end{equation}

\begin{equation}
\mu_{i} = \lambda_{i} + \frac{1}{rt_{i}},
\end{equation}
and
\begin{equation}
\mu_{i}(\phi) = \lambda_{i} + \frac{ln(\frac{1}{1 - \phi})}{rt_{i}(\phi)}
\end{equation}

Equations 2 and 4 follow from the fact that for M/M/1 queues, $P(RT_{i} < rt_{i}(\phi)) = \phi = 1 - e^{-(\mu_{i} - \lambda_{i}rt_{i}(\phi))}$. Without loss of generality, in subsequent sections of this paper, we conduct our analysis on expected steady state parameters. As mentioned previously, due to the Markovian nature of the service process, the case for percentiles is exactly similar to the case for expected steady state analysis, the only difference in analysis being due to the constant, $ln(\frac{1}{1 - \phi})$. Thus, all our proposed equilibrium related results hold true for percentile analysis as well.

Each cloud provider $i$ incurs a fixed cost $c_{i}$ per user request served and a fixed cost $\rho_{i}$ per unit of service capacity provisioned. $c_{i}$ arises due to the factor $\lambda_{i}$ in Equation 3 and $\rho_{i}$ arises due to the factor $\frac{1}{rt_{i}}$ in the same equation. In this sense, our QoS-dependent pricing models are \emph{queueing-driven}. A cloud provider charges $pr_{i}$ to service each end-user request, where $pr_{i}\,\epsilon \, [pr_{i}^{min}, pr_{i}^{max}]$. It is evident that each CP selects a price that results in it accruing a non-negative gross profit margin. The gross profit margin for CP $i$ is given as $pr_{i} - c_{i} - \rho_{i}$, where $c_{i} + \rho_{i}$ is the marginal cost per unit of end-user demand. Thus, the price lower bound, $pr_{i}^{min}$, for each CP $i$ is determined by the following equation.
\begin{equation}
pr_{i}^{min} = c_{i} + \rho_{i}, \, \forall i = 1,...,n
\end{equation}
We define the demand of any CP $i$, $\lambda_{i}$, as a function of the vectors \textbf{pr} = $(pr_{1},.....,pr_{n})$ and \textbf{s} = $(s_{1},......,s_{n})$. Mathematically, we express the demand function as
\begin{equation}
\lambda_{i} = \lambda_{i}(\textbf{pr}, \textbf{s}) = x_{i}(s_{i}) - y_{i}pr_{i} - \sum_{j \ne i}\alpha_{ij}(s_{j}) + \sum_{j \ne i}\beta_{ij}pr_{j},
\end{equation}
where $x_{i}(s_{i})$ is an increasing, concave, and thrice differentiable function in $s_{i}$ satisfying the property of non-increasing marginal returns to scale, i.e., equal-sized reductions in response time results in progressively smaller increases in end-user demand. The functions $\alpha_{ij}$ are assumed to be non-decreasing and differentiable. A typical example of a function fitting $x_{i}(s_{i})$ and $\alpha_{ij}(s_{j})$ is a logarithmic function. We model Equation 6 as a separable function of price and QoS vectors, for ensuring tractable analyses as well as for extracting the independent effects of price and QoS changes on the overall end-user demand. Intuitively, Equation 6 states that QoS improvements by a CP $i$ result in an increase in its end-user demand, whereas QoS improvements by other competitor CPs result in a decrease in its demand. Similarly, a price increase by a CP $i$ results in a decrease in its end-user demand, whereas price increases by other competing CPs result in an increase in its demand. Without loss of practical generality, we also assume 1) a uniform increase in prices by all $n$ CPs cannot result in an increase in any CP's demand volume, and 2) a price increase by a given CP cannot result in an increase in the market's aggregate end-user demand. Mathematically, we represent these two facts by the following two relationships.
\begin{equation}
y_{i} > \sum_{j \ne i}\beta_{ij}, \, i = 1,......,n
\end{equation}
and
\begin{equation}
y_{i} > \sum_{j \ne i}\beta_{ji}, \, i = 1,.......,n
\end{equation}
The long run average profit for CP $i$ in a given time period, assuming that response times are expressed in terms of expected values, is a function of the price and QoS levels of CPs, and is given as
\begin{equation}
P_{i}(\textbf{pr}, \textbf{s}) = \lambda_{i}(pr_{i} - c_{i} - \rho_{i}) - \frac{\rho_{i}}{\overline{rt} - s_{i}},\,\forall i
\end{equation}
The profit function for each CP acts as its \emph{utility/payoff function} when it is involved in price and QoS games with other competing CPs.  We assume in this paper that the profit function for each CP is known to other CPs, but none of the CPs know the values of the parameters that other competing CPs adopt as their strategy. \\ \\
\emph{Problem Statement:} \emph{Given the profit function for each CP (public information), how would each advertise its price and QoS values (without negotiating with other CPs) to end-users so as to maximize its own profit. In other words, in a competitive game of profits played by CPs, is there a situation where each CP is happy with its (price, QoS) advertised pair and does not benefit by a positive or negative deviation in the values of the advertised pair. } \\  

In this paper, we study games involving price and QoS as the primary parameters, i.e., we characterize and analyze the \emph{existence}, \emph{uniqueness}, and \emph{convergence} of Nash equilibria. Our primary goal is to compute the optimal price and QoS levels offered by CPs to its end-users under market competition. Our analysis paves the path for each cloud provider to 1) know what price and QoS levels to set for its clients (end-users) for a given application type, such that it could exist in the cloud market, and 2) practically and dynamically provision appropriate capacity for satisfying advertised QoS guarantees, by taking advantage of the property of \emph{virtualization} in cloud networks. The property of virtualization entails each CP to allocate optimal resources dynamically in a fast manner to service end-user requests. Using our pricing framework, in each time period, cloud providers set the appropriate price and QoS levels after competing in a game; the resulting prices drive end-user demand; the CPs then allocate optimal resources to service demand.

We consider the following types of price-QoS game models in our work.
\begin{enumerate}
\item CP QoS guarantees are pre-specified; CPs compete with each other for prices, given QoS guarantees. (Game 1)
\item CPs compete for price and QoS simultaneously. (Game 2)
\item CP price levels are pre-specified; CPs compete for QoS levels. (Game 3). Game 3 is a special case of Game 2 and in Section 4, we will show that it is a Game 2 derivative. 
\end{enumerate}
\emph{List of Notations:} For reader simplicity, we provide a table of most used notations related to the analysis of games in this paper.

\begin{table}[tb]\tiny
\centering
\begin{tabular}{|c|c|}
\hline
Symbol & Meaning\\
\hline

$U_{i} = P_{i}$ & Utility function of CP $i$\\
$pr_{i}$ & Price charged by CP $i$ per end-user \\
\textbf{pr} & Price vector of CPs\\
$pr^{*}$ & Nash equilibrium price vector\\
$c_{i}$ & Cost incurred by CP $i$ to service each user\\
$\lambda_{i}$ & Arrival rate of end-users to CP $i$\\
$\rho_{i}$ & cost/unit of capacity provisioning by CP $i$\\
$\overline{rt}$ & response time upper bound guarantee\\
$rt_{i}$ & response time guarantee by CP $i$\\
$\phi$ & percentile parameter\\
$s_{i}$ & QoS level guarantee provided by CP $i$ to its users\\
\textbf{s} & QoS vector of CPs\\
$s^{*}$ & Nash equilibrium QoS vector\\
$x_{i}()$ & increasing, concave, and a thrice differentiable function\\
$\alpha_{ij}()$ & non-decreasing and differentiable function\\
\hline
\end{tabular}
\caption{List of Symbols and Their Meaning}
\label{tab:template}
\end{table}

\section{Game 1}
In this section we analyze the game in which the QoS guarantees of CPs are exogenously specified and the CPs compete for prices.\\ \\
\emph{Game Description} \\ \\
\emph{Players:} Individual cloud providers\\
\emph{Game Type:} Non-cooperative, i.e., no interaction between CPs\\
\emph{Strategy Space:} Choosing a price in range $[pr_{i}^{min}, pr_{i}^{max}]$\\
\emph{Player Goal:} To maximize its individual utility $U_{i} = P_{i}$\\

Our first goal is to show that this game has a unique price Nash equilibrium, $pr^{*}$(an instance of vector \textbf{pr}), which satisfies the following first order condition
\begin{equation}
\frac{\partial P_{i}}{\partial pr_{i}} = -y_{i}(pr_{i} - c_{i} - \rho_{i}) + \lambda_{i},\, \forall i,
\end{equation}
which in matrix notation can be represented as
\begin{equation}
\textbf{M}\cdot \textbf{pr} = \overline{\textbf{x}}(\textbf{s}) + \textbf{z},
\end{equation}
where \textbf{M} is an $n \times n$ matrix with $M_{ii} = 2y_{i}$, $M_{ij} = -\beta_{ij},\, i\ne j$, and where $z_{i} = y_{i}(c_{i} + \rho_{i})$.

We have the following theorem and corollary regarding equilibrium results for our game.\\
\textbf{Theorem 1:} \emph{Given that the QoS guarantees of CPs are exogenously specified, the price competition game has a unique Nash equilibrium,} $pr^{*}$\emph{,} \emph{which satisfies Equation 11.} \emph{The Nash equilibrium user demand,} $\lambda_{i}^{*}$\emph{, for each CP} $i$ \emph{evaluates to} $y_{i}(pr_{i}^{*} - c_{i} - \rho_{i})$, \emph{and the Nash equilibrium profits,} $P_{i}^{*}$\emph{, for each CP} $i$ \emph{is given by }$y_{i}(pr_{i}^{*} - c_{i} - \rho_{i})^{2} - \frac{\rho_{i}}{\overline{rt} - s_{i}}$.\\ \\
\emph{Proof:} For a given service level vector \textbf{s}, each CP $i$ reserves a capacity of $\frac{1}{rt_{i}} = \frac{1}{\overrightarrow{rt} - s_{i}}$. Consider the game $G$ with profit/utility functions for each CP $i$ represented as
\begin{equation}
P_{i} = (x_{i}(s_{i}) - y_{i}p_{i} - \sum_{j \ne i}\alpha_{ij}(s_{j}) + \sum_{j \ne i}(\beta_{ij}p_{j})(pr_{i} - c_{i} - \rho_{i}) - W,
\end{equation}
where
\[W = \frac{\rho_{i}}{\overline{rt} - s_{i}}\]
Since $\frac{\partial^{2} P_{i}}{\partial pr_{i}\partial pr_{j}} = \beta_{ij}$, the function $P_{i}$ is \emph{supermodular}\footnote{A function $f: R^{n} \rightarrow R$ is supermodular if it has the following increasing difference property, i.e., $f(m_{i}^{1}, m_{-i}) - f(m_{i}^{2}, m_{-i})$, increases in $m_{i}$ for all $m_{i}^{1} > m_{i}^{2}$ in $(pr_{i}, pr_{j}).$ The readers are referred to \cite{topkis} for more details on supermodularity.}. The strategy set of each CP $i$ lies inside a closed interval and is bounded, i.e., the strategy set is $[pr_{i}^{min}, pr_{i}^{max}]$, which is a \emph{compact} set. Thus, the pricing game between CPs is a \emph{supermodular game} and possesses a Nash equilibrium \cite{vives}. Since $y_{i} > \sum_{j \ne i}\beta_{ij}, \, i = 1,......,n$ (by Equation 7),$ -\frac{\partial^{2} P_{i}}{\partial pr_{i}^{2}} > \sum_{i\ne j}\frac{\partial^{2} P_{i}}{\partial pr_{i}\partial pr_{j}}$ and thus the Nash equilibrium is unique \cite{mr}. Rewriting Equation 11 and using Equation 6, we get $\lambda_{i}^{*} = y_{i}(pr_{i}^{*} - c_{i} - \rho_{i})$. Substituting $\lambda_{i}^{*}$ in Equation 9, we get $P_{i}^{*} = y_{i}(pr_{i}^{*} - c_{i} - \rho_{i})^{2} - \frac{\rho_{i}}{\overline{rt} - s_{i}}$ $\blacksquare$\\ 
\textbf{Corollary 1:} \emph{a)} $pr^{*}$ and $\lambda^{*}$ \emph{are increasing and decreasing respectively in each of the parameters }\{$c_{i}, \rho_{i},\,\,i = 1,2,...,n$\}, \emph{and b)} $\frac{\partial pr_{i}^{*}}{\partial s_{j}} = \frac{1}{y_{i}}\frac{\partial \lambda_{i}^{*}}{\partial s_{j}} = (M^{-1})_{ij}x'_{j}(s_{j}) - \sum_{l \ne j}(M^{-1})_{il}x'_{lj}(s_{j}).$ \\ \\
\emph{Proof:} Since the inverse of matrix \textbf{M}, i.e., $M^{-1}$ exists and is greater than or equal 0\cite{bfr}, from $pr^{*}$ = $M^{-1}(\overline{\textbf{x}}(\textbf{s}) + \textbf{z})$ (Equation 11), we have $pr_{i}^{*}$ is increasing in \{$c_{i}, \rho_{i}\,i = 1,2,...,n$\}. Again, from Lemma 2 in \cite{bfr}, we have $\delta_{i} \equiv y_{i}(M^{-1})_{ii} \Rightarrow 0.5 \le \delta_{i} < 1$, where $\delta_{i}$ is the degree of \emph{positive externality}\footnote{A positive externality is an external benefit on a user not directly involved in
a transaction. In our case, a transaction refers to a CP setting its price and QoS parameters.} faced by CP $i$ from other CP (price, QoS) parameters, and it increases with the $\beta$ coefficients. This leads us to $\frac{\partial pr_{i}}{\partial c_{i}} = \frac{\partial pr_{i}}{\partial \rho_{i}} = y_{i}(M^{-1})_{ii} = \delta_{i} > 0$. Therefore, we show in another different way that $pr^{*}$ is increasing in \{$c_{i}, \rho_{i},\,i = 1,2,...,n$\}. Since $M^{-1}$ exists and is greater than or equal to 0, we again have $\frac{\partial \lambda_{i}}{\partial c_{i}} = \frac{\partial \lambda_{i}}{\partial \rho_{i}} = y_{i}(\frac{\partial pr_{i}}{\partial \rho_{i}} - 1) = y_{i}(\frac{\partial pr_{i}}{\partial c_{i}} - 1) = y_{i}(\delta_{i} - 1) < 0$, from which we conclude that $\lambda^{*}$ is decreasing in \{$c_{i}, \rho_{i},\,i = 1,2,...,n$\}. Part b) of the corollary directly follows from the fact that the inverse of matrix \textbf{M}, i.e., $M^{-1}$ exists, is greater than or equal 0, and every entry of $M^{-1}$ is increasing in $\beta_{ij}$ coefficients. $\blacksquare$

Corollary 1 implies that 1) under a larger value for CP $i's$ degree of positive externality $\delta_{i}$, it is willing to make a bolder price adjustment to an increase in any of its cost parameters, thereby maintaining a larger portion of its original profit margin. The reason is that competing CPs respond with larger price themselves, and 2) there exists a critical value $0 \le s_{ij}^{0} \le \overline{rt}$ such that as CP $j$ increases its QoS level, $pr_{i}^{*}$ and $\lambda_{i}^{*}$ are increasing on the interval $[0, s_{ij}^{0})$, and decreasing in the interval $[s_{ij}^{0}, \overline{rt})$.

\emph{Sensitivity Analysis:} We know the following relationship
\begin{equation}
\frac{\partial P_{i}^{*}}{\partial s_{j}} = 2y_{i}(pr_{i}^{*} - c_{i} - \rho_{i})\frac{\partial pr_{i}^{*}}{\partial s_{j}}
\end{equation}
From it we can infer that CP $i$'s profit increases as a result of QoS level improvement by a competing CP $j$ if and only if the QoS level improvement results in an increase in CP $i$'s price. This happens when $P_{i}^{*}$ increases on the interval $[0, s_{ij}^{0}]$ and decreases on the remaining interval $(s_{ij}^{0}, \overline{rt}]$. In regard to profit variation trends, on its own QoS level improvement, a dominant trend for a CP is not observed. However, we make two observations based on the holding of the following relationship
\begin{equation}
\frac{\partial P_{i}^{*}}{\partial s_{j}} = 2y_{i}(pr_{i}^{*} - c_{i} - \rho_{i})\frac{\partial pr_{i}^{*}}{\partial s_{j}} - \frac{\rho_{i}}{(\overline{rt} - s_{i})^{2}}
\end{equation}
If a CP $i$ increases its QoS level from 0 to a positive value and and this results in its price decrease, $i$'s equilibrium profits become a decreasing function of its QoS level at all times. Thus, in such a case $i$ is better off providing minimal QoS level to its customers. However, when CP $i$'s QoS level increases from 0 to a positive value resulting in an increase in its price charged to customers, there exists a QoS level $s_{i}^{b}$ such that the equilibrium profits alternates arbitrarily between increasing and decreasing in the interval $[0, s_{i}^{b})$, and decreases when $s_{i} \ge s_{i}^{b}$.

\emph{Convergence of Nash Equilibria:} Since the price game in question has a unique and optimal Nash equilibria, it can be easily found by solving the system of first order conditions, $\frac{\partial P_{i}}{\partial pr_{i}} = 0$ for all $i$.
\section{Game 2}
In this section we analyze the game in which the CPs compete for both, price as well as QoS levels. In the process of analyzing Game 2, we also derive Game 3, as a special case of Game 2, and state results pertaining to Game 3. \\ \\
\emph{Game Description} \\ \\
\emph{Players:} Individual cloud providers\\
\emph{Game Type:} Non-cooperative, i.e., no interaction between CPs\\
\emph{Strategy Space:} price in range $[pr_{i}^{min}, pr_{i}^{max}]$ and QoS level $s_{i}$ \\
\emph{Player Goal:} To maximize its individual utility $U_{i} = P_{i}$\\

We have the following theorem regarding equilibrium results.\\
\textbf{Theorem 2:} \emph{Let} $\overline{rt} \le \sqrt[3]{\frac{4\underline{y}\underline{\rho}}{(\overline{x'})^{2}}}$, \emph{where} $\underline{y} = min_{i}\,y_{i}, \underline{\rho} = min_{i}\,\rho_{i},\,\, \overline{x'} = max_{i}\,x_{i}'(0)$. \emph{There exists a Nash equilibrium} $(pr^{*}, s^{*})$, \emph{which satisfies the following system of equations:}
\begin{equation}
\frac{\partial P_{i}}{\partial pr_{i}} = -y_{i}(pr_{i} - c_{i} - \rho_{i}) + \lambda_{i} = 0,\, \forall i,
\end{equation}
\emph{and satisfies the condition that either} $s_{i}(pr_{i})$ \emph{is the unique root of} $x'_{i}(s_{i})(pr_{i} - c_{i} - \rho_{i}) = \frac{\rho_{i}}{(\overline{rt} - s_{i})^{2}}$ if $pr_{i} \ge c_{i} + \rho_{i}(1 + \frac{1}{\overline{rt}^{2}x'_{i}(0)})$ \emph{or} $s_{i}(pr_{i}) = 0$ \emph{otherwise.} \emph{Conversely, any solution of these two equations is a Nash equilibrium. }\\ \\
\emph{Proof:} To prove our theorem, we just need to show that the profit function $P_{i}$ is \emph{jointly concave} in $(pr_{i}, s_{i})$. Then by the \emph{Nash-Debreu} theorem \cite{ft}, we could infer the existence of a Nash equilibria. We know the following results for all CP $i$
\begin{equation}
\frac{\partial P_{i}}{\partial pr_{i}} = -y_{i}(pr_{i} - c_{i} - \rho_{i}) + \lambda_{i}
\end{equation}
and
\begin{equation}
\frac{\partial P_{i}}{\partial \theta_{i}} = x'_{i}(s_{i})(pr_{i} - c_{i} - \rho_{i}) - \frac{\rho_{i}}{(\overline{rt} - s_{i})^{2}}
\end{equation}
Thus, $\frac{\partial^{2}P_{i}}{\partial pr_{i}^{2}} = -2y_{i} < 0$, $\frac{\partial^{2}P_{i}}{\partial s_{i}^{2}} = x''_{i}(s_{i})(pr_{i} - c_{i} - \rho_{i}) - \frac{2\rho_{i}}{(\overline{rt} - s_{i})^{3}} < 0$, $\frac{\partial^{2}P_{i}}{\partial s_{i} \partial pr_{i}} = x'_{i}(s_{i})$. We determine the determinant of the Hessian as $-2y_{i}(x''_{i}(s_{i})(pr_{i} - c_{i} - \rho_{i}) - \frac{\rho_{i}}{(\overline{rt} - s_{i})^{2}} \ge 0$ (the sufficient condition for $P_{i}$ to be jointly concave in $(pr_{i}, s_{i}))$, if the following condition holds:
\begin{equation}
\frac{4y_{i}\rho_{i}}{\partial pr_{i}^{2}} \ge (x'_{i}(s_{i}))^{2} \Leftrightarrow \overline{rt} \le min_{s_{i}}\,\sqrt[3]{\frac{4y_{i}\rho_{i}}{(x'_{i}(s_{i}))^{2}}} = \sqrt[3]{\frac{4y_{i}\rho_{i}}{(x'_{i}(0))^{2}}},
\end{equation}
where the last equality follows from the fact that $x'_{i} > 0$ and $x'_{i}$ is decreasing. Now since $pr^{*} = pr^{*}(s^{*})$, by Theorem 1 it is in the closed and bounded interval $[pr^{min}, pr^{max}]$ and must therefore satisfy Equation 15. Again from Equation 17, we have $\frac{\partial P_{i}}{\partial s_{i}} \rightarrow -\infty$ as $s_{i}$ tends to $\overline{rt}$, which leads us to the conclusion that $s_{i}(pr_{i})$ is the unique root of $x'_{i}(s_{i})(pr_{i} - c_{i} - \rho_{i}) = \frac{\rho_{i}}{(\overline{rt} - s_{i})^{2}}$ if $pr_{i} \ge c_{i} + \rho_{i}(1 + \frac{1}{\overline{rt}^{2}x'_{i}(0)})$ or $s_{i}(pr_{i}) = 0$ otherwise. $\blacksquare$

\emph{Sensitivity Analysis:} We know that $s_{i}(pr_{i})$ depends on $x'_{i}(s_{i})$ and $pr_{i}$. Thus, from the \emph{implicit function theorem} \cite{rudin} we infer that the QoS level of CP $i$ increases with the increase in its Nash equilibrium price. We have the following relationship for $pr_{i} > c_{i} + \rho_{i}(1 + \frac{1}{\overline{rt}^{2}x'_{i}(0)})$,
\begin{equation}
s'_{i}(pr_{i}) = \frac{x'_{i}(s_{i})}{x''_{i}(s_{i})(pr_{i} - c_{i} - \rho_{i}) - \frac{\rho_{i}}{(\overline{rt} - s_{i})^{2}}} > 0,
\end{equation}
whereas $s'_{i}(pr_{i}) = 0$ for $pr_{i} < c_{i} + \rho_{i}(1 + \frac{1}{\overline{rt}^{2}x'_{i}(0)})$. We also notice that for $pr_{i} > c_{i} + \rho_{i}(1 + \frac{1}{\overline{rt}^{2}x'_{i}(0)})$, $s_{i}^{*}$ increases concavely with $pr_{i}^{*}$. The value of $s_{i}(p_{i})$ obtained from the solution of the equation $x'_{i}(s_{i})(pr_{i} - c_{i} - \rho_{i}) = \frac{\rho_{i}}{(\overline{rt} - s_{i})^{2}}$ if $pr_{i} \ge c_{i} + \rho_{i}(1 + \frac{1}{\overline{rt}^{2}x'_{i}(0)})$, can be fed into Equation 15 to compute the price vector. The system of equations that result after substitution is \emph{non-linear} in vector \textbf{pr} and could have multiple solutions, i.e., multiple Nash equilibria.\\ \\
\emph{Inferences from Sensitivity Analysis: Games 1, 2, and 3 gives us non-intuitive insights to the price-Qos changes by individual CPs. We observe that the obvious intuitions of equilibrium price decrease of competing CPs with increasing QoS levels and vice-versa do not hold under all situations and sensitivity analysis provide the conditions under which the counter-result holds. Thus, the intricate nature of non-cooperative strategy selection by individual CPs and the interdependencies of individual strategies on the cloud market make cloud economics problems interesting.} \\ 

\emph{Convergence of Nash Equilibria:} Since multiple Nash equilibria might exist for the price vectors for the simultaneous price-QoS game, the \emph{tatonnement scheme} \cite{hv}\cite{arrow} can be used to prove convergence. This scheme is an iterative procedure that numerically verifies whether multiple price equilibria exist, and uniqueness is guaranteed if and only if the procedure converges to the same limit when initial values are set at $pr^{min}$ or $pr^{max}$. Once the equilibrium price vectors are determined, the equilibrium service levels are easily computed. If multiple equilibria exist the cloud providers select the price equilibria that is component-wise the largest.

Regarding the case when CP price vector is given, we have the following corollary from the result of Theorem 2, which leads us to equilibrium results of \emph{Game 3}, a special case of Game 2.\\
\textbf{Corollary 2.} \emph{Given any CP price vector,} $pr^{f}$\emph{, the Nash equilibrium} $s(pr^{f})$ \emph{is the dominant solution in the QoS level game between CPs, i.e., a CP's equilibrium QoS level is independent of any of its competitors cost or demand characteristics and prices. When} $s_{i}(pr^{f}) > 0$\emph{, the equilibrium QoS level is increasing and concave in} $pr_{i}^{f}$\emph{, with} $s'_{i}(pr_{i}^{f}) = \frac{-x'_{i}(s_{i})}{x''_{i}(s_{i})(pr_{i}^{f} - c_{i} - \rho_{i}) - \frac{2\rho_{i}}{(\overline{rt} - s_{i})^{3}}}$.\\ \\
\emph{Proof:} Substituting $pr^{max} = pr^{min} = pr^{f}$ into Theorem 2 leads us to the fact that $s(pr^{f})$ is a Nash equilibrium of the QoS level competition game amongst CPs and that it is also a \emph{unique} and a \emph{dominant} solution, since $s(pr^{f})$ is a function of $pr_{i}$, $c_{i}$, and $\rho_{i}$ only. (Following from the fact that $s_{i}(pr_{i})$ is the unique root of $x'_{i}(s_{i})(pr_{i} - c_{i} - \rho_{i}) = \frac{\rho_{i}}{(\overline{rt} - s_{i})^{2}}$ if $pr_{i} \ge c_{i} + \rho_{i}(1 + \frac{1}{\overline{rt}^{2}x'_{i}(0)})$ or $s_{i}(pr_{i}) = 0$ otherwise.) $\blacksquare$

We observe that Game 3 being a special case of Game 2 entails a unique Nash equilibrium, whereas Game 2 entails multiple Nash equilibria. 

\section{Conclusion and Future Work}
In this paper, we developed inter-organizatinal economic models for pricing cloud network services when several cloud providers co-exist in a market, servicing a single application type. We devised and analyzed three price-QoS game-theoretic models relevant to cloud networks. We proved that a \emph{unique} pure strategy Nash equilibrium (NE) exists in two of our three QoS-driven pricing models. In addition, we also showed that the NE's \emph{converge}; i.e., there is a \emph{practically implementable} algorithm for each model that computes the NE/s for the corresponding model. Thus, even if no unique Nash equilibrium exists in some of the models, we are guaranteed to find the largest equilibria (preferred by the CPs) through our algorithm. 

Our price-QoS models can drive optimal resource provisioning in cloud networks. The NE price and QoS levels for each cloud provider drives optimal end-user demand in a given time period w.r.t. maximizing individual CP profits under competition. Servicing end-user demands requires provisioning capacity. As a part of future work, we plan to extend our work to develop queueing optimization models to compute optimal provisioned resources in cloud networks. Once the optimal values are computed, the power of virtualization in cloud networks makes it possible to execute dynamic resource provisioning in a fast and efficient manner in multiple time periods. Thus, our pricing models are specifically suited to cloud networks. As a part of future work, we also plan to extend our analysis to the case where cloud providers are in simultaneous competition with other CPs on multiple application types.

%\[\textbf{S}^{*} = argmax_{\textbf{S}^{*}\,\epsilon\,\textbf{U}}\, NP(\textbf{S}) = \prod_{i = 1}^{n}S_{i}^{\beta_{i}}\]
%\[s.t \sum_{i = 1}^{n}\frac{\theta_{i}S_{i}}{k - D_{0i}S_{i}} = B_{max} - \sum_{i = 1}^{n}R_{0i}\]
%\[S_{i} > 0\,\forall i;\, \sum_{i = 1}^{n}\beta_{i} = 1\]

\bibliography{alluvion}
\bibliographystyle{plain}
%H.~Kopka and P.~W. Daly, \emph{A Guide to {\LaTeX}}, 3rd~ed.\hskip 1em plus

%\end{thebibliography}

\end{document}